\documentstyle[12pt]{article}
\begin{document}
\newcommand{\be}{\begin{equation}}
\newcommand{\ee}{\end{equation}}

\begin{center}

{\bf Cherenkov radiation and pair production\\
by particles traversing laser beams}\\

\vspace{3mm}

I.M. Dremin\\

Lebedev Physical Institute, Moscow 119991, Russia\\ 

\end{center}

\begin{abstract}
It is shown that Cherenkov radiation can be observed at TESLA in electron
collisions with optical laser pulses. The prospects for it to be observed at
SLC, LEP, LHC and RHIC are discussed. The conclusions are compared with results
for pair production.
\end{abstract}

The problem of very high energy charged particles collisions with laser beams is
widely discussed now, mostly in connection with the $e^+e^-$-pair production
(see latest references \cite{taji, ring, ahrs, popo}) at SLC and TESLA X-ray
laser facilities \cite{arth, bmrw, mtsc} and with some other issues of
fundamental physics.

Spontaneous particle creation from vacuum induced by a strong external field was
theoretically considered in many papers beginning with Refs \cite{saut, heul, schw}.
However, very powerful high-frequency lasers are needed for this process to be
observed. First observations of this effect were done at SLC \cite{bula, burk}.

Here, I would like to note that optical lasers can be used for studies of 
Cherenkov radiation. It is crucial for its observation that the main background
process of Compton scattering does not contribute to the kinematical region of
Cherenkov radiation.
The principal possibility of such a process was first mentioned by V. Ritus
in Ref. \cite{ritu}. The results can be, in general, applied to verify our
ideas about the properties of the "photon medium" in the region where new
physics concepts can become essential and to measurements of beam
energy and laser bunch parameters.

X-ray lasers have been proposed for use in $e^+e^-$-pair
production studies because the quanta energies are high enough to reach
the threshold energy which in c.m.s. is equal to $m+2m_e$, where $m_e$ is
the electron mass and $m$ is the accelerated particle mass (equal to $m_e$
at SLC, LEP, TESLA, to the proton mass at LHC and to the nucleus mass at RHIC).
At the same time, for studies of Cherenkov radiation other characteristics of
a laser, namely, the ratio $F/F_0$ of the electric field $F$ to its "critical"
value $F_0=m_e^2/e$ or, equivalently, its peak  power density $S$, are
important. They determine the index of refraction $n$ of
the "photon medium" in laser bunches.
The difference of $n$ from 1 is proportional to the density of photons in a
laser pulse, i.e., to $S$. Just this difference defines the threshold of
Cherenkov radiation, its emission angle and intensity\footnote{The analogous
problem was considered in Ref. \cite{drem} for particles traversing the
cosmic microwave background radiation. The density of relic photons is, however,
extremely low and, therefore, the index of refraction is so close to 1 that the
threshold energy is too high for this effect to be observable.}. The parameters
$F/F_0$ and $S$ are higher for optical lasers than for presently available
X-ray lasers.
That is why they would be preferred for Cherenkov radiation studies nowadays.
Besides, the energy limitations also favour optical lasers for this purpose.

The necessary conditions for Cherenkov radiation to be observed are the excess
of the index of refraction $n$ over 1, i.e.
\be
\Delta n=n-1>0      \label{delt}
\ee
and the real emission angle, given by the formula 
\be
\cos \theta=\frac {1}{\beta n},     \label{cost}
\ee
where $\beta =v/c=\sqrt {1-\frac {m^2}{E^2}},\;  m, E$ are the particle mass and
energy. For small values of $m/E$ and $\Delta n$ one gets
\be
\theta \approx \sqrt {2\Delta n - \frac {m^2}{E^2}}=
\sqrt {2\Delta n-\gamma ^{-2}}.  \label{thet}
\ee
Hence, the condition for the energy to exceed the threshold for Cherenkov
radiation $E_{Ct}$ is written as
\be
\gamma m=E\geq E_{Ct}=\frac {m}{\sqrt {2\Delta n}}=\gamma _{Ct}m.   \label{ethr}
\ee
It is easily seen that the threshold can become very high for small $\Delta n$.

The formula (\ref{thet}) can be rewritten as
\be
0\leq \theta =\frac {\sqrt {\gamma ^2-\gamma _{Ct}^2}}{\gamma \gamma _{Ct}}\leq
\frac {1}{\gamma _{Ct}}=\theta _{max} \;\;\; (\gamma \rightarrow \infty ).
\label{thga}
\ee
It is seen that the emission angles of Cherenkov radiation 
increase from 0 at the threshold to $\theta _{max}=\gamma _{Ct}^{-1}$ for
$\gamma \rightarrow \infty $. However, already at $E=2E_{Ct}$ this angle is
very close to $\theta _{max}$ ($\theta (2E_{Ct})\approx 0.866\; \theta _{max}$).

The number of Cherenkov photons emitted by a single particle with the electric
charge $e$ in the interval of frequencies $d\omega $ from the path length $dl$
is given by the common expression \cite{tfra}
\be
\frac {dN_1}{d\omega dl}=2\alpha \Delta n,     \label{adn}
\ee
where the fine structure constant $\alpha \approx 1/137$.
Thus all physical characteristics of the process are determined by the value
$\Delta n$. The intensity of the radiation (\ref{adn}) decreases with the
threshold energy (\ref{ethr}) increase:
\be
\frac {dN_1}{d\omega dl}=\frac {\alpha m^2}{E_{Ct}^2}=
\frac {\alpha }{\gamma _{Ct}^2} .     \label{adn1}
\ee

The value of $\Delta n$ is uniquely related to the polarization operator of
$\gamma -\gamma $ scattering. For high energy electrons (protons), the laser
field can be considered as the constant crossed (or null) field. The
refractivity index can be expressed in terms of the photon mass acquired
in such a field. It has been calculated in Refs \cite{naro, ritu} and its
graphical representation can be found in Refs \cite{ritu, rit1} (for other approaches
see also Refs \cite{fmmi, mais, thom}). According to the results of Refs
\cite{naro, ritu}, the value of $\Delta n$ depends on the photon mass $\mu $
and its energy $\omega $ in a following way
\be
\Delta n=-{\rm Re}\mu ^2/2\omega ^2.        \label{deln}
\ee
In head-on collisions, the photon mass depends only on the invariant variable 
\be
\kappa =\frac {2\omega }{m_e}\cdot \frac {F}{F_0}.    \label{kapp}
\ee
The value of Re$\mu ^2$ is negative\footnote{We consider the value for
transverse polarized photons. For other polarizations it differs by a factor
less than 2, and this does not change general conclusions.} in the region about
\be
0\leq \kappa \leq 15 \label{0k15}
\ee
and has a minimum at $\kappa \approx 5$ with
${\rm Re}\mu ^2\approx -0.2\alpha m_e^2$. According to eq. (\ref{deln}), the
refractivity index exceeds 1 in this region, and, consequently, the Cherenkov
radiation is possible at these values of $\kappa $. The perturbation theory is
still applicable \cite{ritu} because $\alpha \kappa ^{2/3}\ll 1$.

At a fixed laser intensity, i.e., a fixed ratio $F/F_0$, the index of refraction
does not depend on $\omega $ at low energies
\be
\Delta n\approx \frac{14\alpha F^2}{45\pi F_0^2},    \label{dn}
\ee
because Re$\mu ^2$ is proportional to $\kappa ^2$ at small $\kappa ^2$.
Thus, the ratio $F/F_0$ defines there main features of Cherenkov radiation.

The Cherenkov threshold $\gamma _{Ct}$ is also completely determined by this
ratio as seen from formulas (\ref{ethr}), (\ref{dn}). It is the same for
electrons and protons. Therefore, the threshold energies $E_{Ct}$ are
approximately 2000 times higher for protons than for electrons. The formulas
(\ref{thet}) and (\ref{dn}) show that, in principle, by measuring the angle
$\theta $ one can get the energy of the particle beam and the strength of
the laser field or its peak power density.

The magnitude of $\Delta n$ decreases at higher values of $\kappa $ and becomes
negative at $\kappa >15$ so that Cherenkov radiation is impossible there.

Even though it can again become positive at extreme energies where the hadronic
channels are important, this region is completely inaccessible in collisions
with laser beams.

Let us remind that the energy threshold for the $e^+e^-$-pair production processes
in high energy head-on collisions of a particle of mass $m$ with laser quanta
is given by
\be
E_{th}\approx \frac {m_e(m+m_e)}{\omega _L}.    \label{ept}
\ee
It depends on the energy of laser quanta $\omega _L$ and is
much lower for X-ray lasers than for optical lasers. It is
approximately 1000 times higher for protons than for electrons. In particular,
the threshold for $\gamma \gamma $-collisions follows from (\ref{ept}) at $m=0$.

The condition for Cherenkov radiation threshold to be below the pair
production threshold imposes the restriction on the laser quanta energies:
\be
\omega_L<\sqrt{2\Delta n}m_e(1+\frac {m_e}{m})=
\gamma _{Ct}^{-1}m_e(1+\frac {m_e}{m}).     \label{omeg}
\ee
The condition (\ref{omeg}) differs for electron and proton
beams only by a factor about 2 in the right hand side. 

For optical and X-ray lasers, according
to \cite{ring}, we accept, correspondingly, $\omega _L=1.2$ eV (actually, it
varies from 0.12 eV for CO$_2$-laser to 2.35 eV for Nd:glass laser) and
3.1 keV, the ratios $F/F_0=3\cdot 10^{-4}$ and $10^{-5}$ or ,equivalently,
the peak power densities $S=3\cdot 10^{22}$W/cm$^2$=$5\cdot 10^{16}$
eV$^4$ and $8\cdot 10^{19}$ W/cm$^2$ (with a possible goal $7\cdot 10^{29}$
W/cm$^2$).

At these parameters, the laser field can be treated as a constant crossed
(or null) field because its invariants (see \cite{ritu})
\be
x=\frac {m_e}{\omega _L}\cdot \frac {F}{F_0}; \;\;\;\; \chi =2\gamma \frac{F}{F_0}
\label{xchi}
\ee
are large. For optical lasers with $F/F_0=3\cdot 10^{-4}$, one gets $x=125$,
and at TESLA energies $\gamma =10^6$, $\chi =600$. Thus, the wavelength 
$1/\omega _L$ is much larger than the formation length $m_e/eF$.

Electromagnetic processes at these conditions are extremely interesting.
New physics concepts can be necessary here because the effective expansion
parameter \cite{mnar, nar1, rit1} $\alpha \chi ^{2/3}$ exceeds 0.5. In 
particular, this would indicate that the constant field allows the interaction
with field quanta of the arbitrarily low energies. Radiation effects should be
reconsidered.

At $F/F_0=3\cdot 10^{-4}$, the relations (\ref{kapp}), (\ref{0k15}) impose the
upper limit $\omega <12$ GeV. Only this region of comparatively low energies
is admissible for Cherenkov quanta in such strong fields.

Using these characteristics, one also concludes that the condition
(\ref{omeg}) is satisfied for optical lasers with $F/F_0>3\cdot 10^{-5}$
($S>3\cdot 10^{21}$ W/cm$^2$) and not valid for
the presently available X-ray lasers. To satisfy it for X-ray lasers, one must
achieve the peak power density as high as 10$^{27}$ W/cm$^2$ which is,
nevertheless, within the proclaimed goals \cite{ring}.

It follows from eqs (\ref{adn}) and (\ref{dn}) that one should deal with most 
intensive laser fields to get higher intensity of Cherenkov radiation. Thus,
in what follows, we discuss only optical lasers briefly referring to X-ray
lasers for some estimates.

The numerical value of $\Delta n$ for $\gamma $-quanta in the optical laser
field with $F/F_0=3\cdot 10^{-4}$ is given by
\be
\Delta n=0.65\cdot 10^{-10}.   \label{dnn}
\ee
Therefore, the typical angles and threshold $\gamma $-factors for Cherenkov
radiation are
\be
\theta _{max}=1.14\cdot 10^{-5}; \;\;\;\; \gamma _{Ct}=8.8\cdot 10^4. \label{thg}
\ee
This implies that the energy threshold for Cherenkov radiation is exceeded at
LEP2 and TESLA since it is $E_{Ct}^{(e)}=45$ GeV and is close to the upper
energy of SLC. Only with further increase of the laser power, it would be
possible to study this process at SLC.

The pair production threshold for optical lasers is about 430 GeV. Thus SLC and
LEP energies are well below it while TESLA is just close\footnote{However, notice
the rather wide spread of available wavelengths for optical lasers mentioned
above.} to the threshold value.

For proton beams the Cherenkov radiation
threshold $E_{Ct}^{(p)}=83$ TeV is too high even for LHC. If the optical
lasers with $F/F_0 > 4\cdot 10^{-3}$ 
(the peak power density $S > 5\cdot 10^{24}$ W/cm$^2$) will become
accessible, one can hope to observe this effect there as well. This energy
is much lower than the threshold for pair production at proton accelerators
which is about 400 TeV.

What concerns the X-ray laser facilities, the threshold for pair production
(\ref{ept}) is well below the energies accessible at all high energy accelerators.

Now, let us calculate the intensity of the Cherenkov radiation for electron
beams and compare it with the main background process of Compton
scattering\footnote{The final results are valid for any charged particles.}. 
The total number of Cherenkov quanta emitted in the energy interval $d\omega $
by a particle which collides with the
laser bunch of the coherent spike length $L$ is
\be
\frac {dN_{Ch}}{d\omega }=2\alpha \Delta nL=
1.1\cdot 10^{-5}L\frac {F^2}{F_0^2}. \label{nche}
\ee
This is the energy distribution at low energies as given by eqs (\ref{adn}),
(\ref{deln}). It is almost constant at low energies as demonstrated by eq.
(\ref{nche}) but should decrease towards the cut-off at $\kappa \approx 15$. 
For a fixed value of the ratio $F/F_0$, the energies of emitted Cherenkov
quanta are proportional to $\kappa $ and limited according to eqs (\ref{kapp}),
(\ref{0k15}). However, already at $\kappa  \approx 4$ the magnitude of
$\Delta n$ and, consequently, the intensity (\ref{adn}) are about twice lower
than at $\kappa =0$. Therefore, the effective values of $\omega $ important in
the distribution can be approximately estimated according to eq. (\ref{kapp}) as
\be
\omega _{eff}<2m_e\frac {F_0}{F}.
\ee
For the values of the ratio $F/F_0$ adopted above, one gets
$\omega _{eff}^{(o)}<3.5$ GeV for optical lasers and
$\omega _{eff}^{(X)}<100$ GeV for X-ray lasers. One can use eq. (\ref{kapp})
for an estimate of $\Delta n$ in these energy regions. The threshold and
effective energy of Cherenkov quanta decrease while the emission angle and
the intensity of radiation increase with increase of laser fields $F$.

The absolute intensity can be evaluated according to the formulas (\ref{adn}),
(\ref{nche}). For the coherent spike length $L\sim 1$ mm, the number of quanta
per 1 GeV is estimated as
\be
\frac {dN_{Ch}}{d\omega }\approx 5 \;  {\rm GeV}^{-1}.  \label{nfn}
\ee
Thus the emitted energy within the effective interval should be of the order
of 30 GeV per 1 mm.

To proceed with similar estimates for Compton scattering, we consider first
its kinematics. This leads to the following relation between the emission angle
$\theta $ and energy $\omega $ of the scattered quantum in the laboratory system:
\be
\cos \theta =\left(1+\frac {\omega _L}{E}-\frac {2\omega _L}{\omega }\right)
\left(1-\frac 
{\omega _L}{E}\right)^{-1}\approx 1-\frac {2\omega _L}{\omega }.   \label{thco}
\ee
The precise limits imposed by this relation on the
energy of emitted quanta are given by $\omega _L\leq \omega \leq E$. 
In the right hand side of (\ref{thco}) we have taken into account that
the particle beam energy is much higher than the photons energies
$E\gg \omega > \omega _L$. At the angles typical for Cherenkov radiation
(\ref{thg}) the energy of the backscattered quantum obtained from eq.
(\ref{thco}) is equal to $\omega \approx 37$ GeV while Cherenkov radiation
is much softer ($\omega _{Ch}<3.5$ GeV) due to the cut-off imposed by the
behaviour of $\Delta n$. Such "soft" photons are emitted at
larger angles at Compton scattering. Therefore, there is no overlap of the 
kinematical regions available for Compton and Cherenkov processes. By
separating the relatively soft quanta at the angle (\ref{thg}) one would be able
to get rid of the background due to Compton processes. These processes
contribute to completely different energy ranges and, therefore, can be easily
disentangled.

The multiphoton processes would lead to even harder quanta in Compton
scattering. In other words, they are effective at larger
angles $\theta >2\sqrt {\omega _L/\omega }$. Thus, even at high values of $x$
usual Compton effect formulas are needed to estimate its intensity at small
angles. The non-linear effects of the laser field are not important at Cherenkov
angles (\ref{thg}). Let us also note that these energies are much lower than the
threshold for pair production (\ref{ept}) equal to 220 GeV for $\omega _L=1.2$
eV.

The total energy loss due to Compton scattering is much larger than for Cherenkov
radiation  (\ref{nfn}). It can be as large as 16/63 of the initial electron
energy \cite{ritu}. This is determined by the extremely hard backscattered quanta.

The use of different lasers inevitably leads 
to the change of corresponding values of $F/F_0$ and $S$, and, consequently,
of the threshold energy $E_{Ct}$. Therefore, one should be careful in
estimates. The practical feasibility of
observation of such an effect at high energies should be considered in close
relation to the definite conditions of a particular experiment.
For example, one can not use the Nd:glass laser of SLC experiments
\cite{bula, burk, meli} with $F/F_0=2.3\cdot 10^{-6}$ at
TESLA because the Cherenkov threshold energy becomes too high.
The optimum choice would be the laser with highest power density (see
(\ref{nche})).

At the end, let us note that, in principle, the heavy ion accelerator RHIC
can be also used for pair production studies with X-ray lasers because the
threshold energy according to (\ref{ept}) is equal to 165 GeV per nucleon
that is available there. The Cherenkov radiation threshold is the same as
for proton accelerators if estimated per nucleon. It has been discussed above
in connection with LHC and is not reachable at RHIC. 

I am grateful to E.L. Feinberg, V.A Maisheev, A.I. Nikishov and V.I. Ritus for
discussions and comments.

This work is supported by the RFBR grant 00-02-16101.

\end{document}